\documentclass[aps,prb,twocolumn,showpacs,preprintnumbers]{revtex4}
\usepackage{graphicx}
\begin{document}

%\preprint{submitted to Phys. Rev. B}

\title{Surface losses and self-pumping effects in a long\\
Josephson junction --- a semi-analytical approach}
\author{Marek Jaworski}
\affiliation{Institute of Physics, Polish Academy of Sciences, Al.
Lotnik\'ow 32/46, 02-668 Warszawa, Poland}

\begin{abstract}
The flux-flow dynamics in a long Josephson junction is studied both
analytically and numerically. A realistic model of the junction is
considered by taking into account a nonuniform current distribution,
surface losses and self-pumping effects. An approximate analytical
solution of the modified sine-Gordon equation is derived in the form
of a unidirectional dense fluxon train accompanied by two oppositely
directed plasma waves. Next, some macroscopic time-averaged
quantities are calculated making possible to evaluate the
current-voltage characteristic of the junction. The results obtained
by the present method are compared with direct numerical simulations
both for the current-voltage characteristics and for the loss factor
modulated spatially due to the self-pumping. The comparison shows
very good agreement for typical junction parameters but indicates
also some limitations of the method.
\end{abstract}

 \maketitle

\section{Introduction}

In recent years the flux-flow (FF) dynamics in a long Josephson
junction has attracted considerable attention in view of possible
applications in superconducting mm-wave electronics
\cite{jpn,kosh1,kosh2,kosh3}. The FF mode appears in a junction
immersed in a sufficiently large external magnetic field and can be
described briefly as a unidirectional viscous flow of a dense train
of fluxons (magnetic flux quanta).

Recently, the FF oscillators have found various applications, e.g.
in mm-wave integrated receivers \cite{kosh3,kosh4,kosh5,kosh6},
however some important parameters, such as the radiation line-width
are still not satisfactory. Therefore, there is still a need for
more adequate description of a real Josephson junction operating in
the FF mode, by taking into account some additional factors which
may affect the current-voltage ($I$-$V$) characteristic of the
junction and change its working conditions.

In the majority of papers dealing with long Josephson junctions (see
e.g. Refs.~\onlinecite{sml,lom,km,cir,ss1,mj1,mj2}) a simplified
version of the the sine-Gordon (sG) equation is considered, usually
neglecting surface losses and assuming uniform distribution of the
bias current density. Only recently, a few papers have been
published \cite{pank1,pank2,pank3,mj3} taking into account a
nonuniform bias current distribution and its influence on the
junction behavior. Moreover, in Refs.~\onlinecite{pank2,pank3} a
realistic model of the FF oscillator has been investigated both
experimentally and numerically, including general boundary
conditions, nonuniform bias current profile, surface losses as well
as self-pumping effects \cite{tf} related to additional tunneling of
quasiparticles due to the Josephson radiation.

The aim of the present paper is to present an analytical approach to
the modified sG equation, which takes into account:(i) nonuniform
current distribution, (ii) surface losses, and (iii) spatial
modulation of the loss factor resulting from the self-pumping
effect. Contrary to Ref.~\onlinecite{pank2} we assume standard
open-circuit boundary conditions to make the influence of various
effects more pronounced. Nevertheless, the present analysis can be
easily extended to include also more general boundary conditions.

In the particular case of uniform current distribution, the present
method makes it possible to obtain fully analytical closed-form
expressions describing both the superconducting phase within the
junction and the $I$-$V$ characteristic. Such a solution can be
regarded as the first-order approximation which appears sufficiently
accurate for some moderate junction parameters. However, in the
general case, particularly for very long and weakly damped
junctions, such an analytical approximation is only the first step
in an iterative procedure, which has to be performed numerically.
Thus, in spite of analytical expressions describing the
superconducting phase within the junction, the present approach has
been named ``semi-analytical''.

The paper is organized as follows. In Sec.~II we formulate the
problem, i.e. we present a generalized sG equation subject to
open-circuit boundary conditions at the junction ends. Approximate
analytical solutions to the sG equation are discussed in Sec.~III.
We start with a linearized (small-amplitude) solution and apply
appropriate boundary conditions. Next, some large-amplitude
corrections are introduced, we discuss also possible self-pumping
effects and their influence on the $I$-$V$ characteristic. In
Sec.~IV analytical results are compared with direct numerical
simulations. In particular, we discuss the influence of surface
losses both for junctions of moderate length and for more realistic
structures, such as very long junctions with small damping and
strongly nonuniform current distribution. Sec.~V contains concluding
remarks, we indicate also possible extensions of the method by
taking into account more general boundary conditions.

\section{Formulation of the problem}

Fluxon dynamics in a long Josephson junction is usually described by
the following modified sG equation \cite{sml,lom,km,cir,ss1}:
\begin{equation}
\label{sg1} \phi_{xx}-\phi_{tt}-\alpha\phi_t = \sin\phi-\gamma,
\end{equation}
where $\phi$ denote the quantum phase difference across the barrier,
$\alpha$ is the loss factor, and $\gamma$ is the bias current
density. The spatial coordinate $x$ has been normalized to the
Josephson penetration depth $\lambda_J$ and the time coordinate $t$
to the inverse plasma frequency $\omega_0^{-1}$, where
$\lambda_J=(\hbar/2\mu_0edj_c)^{1/2}$, $\omega_0=(2ej_c/\hbar
C)^{1/2}$, $j_c$ is the critical current density and $C$ denotes the
junction capacitance per unit area.

However, to describe more adequately a real physical situation, one
can consider a more general form \cite{pank2,pank3}:
\begin{equation}
\label{sg2} \phi_{xx}-\phi_{tt}-\alpha(x)\phi_t+\beta\phi_{xxt} =
\sin\phi-\gamma(x),
\end{equation}
where $\beta$ denotes the surface loss parameter and we assume both
$\alpha$ and $\gamma$ to be $x$-dependent, taking into account both
a spatial modulation of the loss factor due to the self-pumping and
a nonuniform distribution of the current density along the junction.

For the overlap geometry one can neglect the self-fields and assume
simple open-circuit boundary conditions \cite{lom}:
\begin{equation}
\label{bc1} \phi_x(\pm L/2)+\beta\phi_{xt}(\pm L/2) = h,
\end{equation}
where $h=H_{\rm ext}/j_c\lambda_J$, $H_{\rm ext}$ denotes the
external magnetic field, and $L\gg 1$ is the normalized junction
length.

Following Refs. \onlinecite{ss1,cir,mj1,mj2,mj3} we look for an
approximate solution in a form of a dense fluxon train traveling on
a rotating background. Thus, a linearized solution of
Eq.~(\ref{sg2}) can be written as
\begin{equation}
\label{sol1} \phi=\phi_0+\psi,
\end{equation}
where $\phi_0=\theta(x)+\Omega t$ is the background term (linear in
time) and $\psi$ denotes a quasi-linear term (usually small)
representing the motion of fluxons and plasma waves within the
junction.

Substituting Eq.~(\ref{sol1}) into Eq.~(\ref{sg2}) we find
\begin{equation}
\label{eq1} \theta_{xx} + \psi_{xx} - \psi_{tt} - \alpha\Omega -
\alpha\psi_t+\beta\psi_{xxt} = \sin(\phi_0 + \psi) - \gamma.
\end{equation}

We are interested in a steady-state, time periodic solution, thus
the background frequency $\Omega$ should be equal to the fundamental
frequency of the oscillating term $\psi$. Both experimental data and
numerical simulations show that the output signal of a real FF
oscillator is nearly sinusoidal \cite{jpn,pank1}. Thus, it is
reasonable to assume the time-dependence of the term $\psi$ to be
harmonic in $\Omega t$, and neglect any higher harmonics generated
by the nonlinear term $\sin(\phi_0+\psi)$. Consequently,
Eq.~(\ref{eq1}) can be split into a time-independent part and a part
oscillating with frequency $\Omega$, while the boundary conditions
(\ref{bc1}) can be written separately for $\theta(x)$ and
$\psi(x,t)$:
\begin{equation}
\label{bc2} \theta_x(\pm L/2)= h,
\end{equation}
\begin{equation}
\label{bc3} \psi_x(\pm L/2)+\beta\psi_{xt}(\pm L/2) = 0.
\end{equation}

\section{Approximate analytical solution}

The method for solving Eq.~(\ref{eq1}) with the boundary conditions
(\ref{bc2}) and (\ref{bc3}) is similar to that reported recently
(see Ref. \onlinecite{mj3} for details). Thus, below we discuss
briefly only the main points of the analysis.

\subsection{Time-independent equation}

It can be easily shown that for $\psi$ oscillating with frequency
$\Omega$, the nonlinear term $\sin(\phi_0+\psi)$ contributes also to
the time-independent part of Eq.~(\ref{eq1}), and such a
contribution yields, in fact, the first-order approximation of the
$I$-$V$ characteristic of the junction. Thus, the time-independent
equation can be written as:
\begin{equation}
\label{theta} \theta_{xx}=\alpha(x)\Omega+S(x)-\gamma(x),
\end{equation}
where
\begin{equation}
\label{s} S(x)=\langle\sin(\phi_0+\psi)\rangle_T= (1/T)\int_0^T
\sin(\phi_0+\psi)dt,
\end{equation}
and  $T=2\pi/\Omega$.

Solving Eq.~(\ref{theta}) for $\theta(x)$ we find
\begin{equation}
\label{soltheta}
\theta=\int_{-L/2}^x\left[\int_{-L/2}^\eta[\alpha(\xi)\Omega+S(\xi)-\gamma(\xi)]
d\xi\right]d\eta+C_1x+C_2.
\end{equation}

Without loss of generality $C_2$ can be set to zero while from the
boundary condition (\ref{bc2}) we find $C_1=h$ and
\begin{equation}
\label{iv}
\gamma_0=\alpha_0\Omega+\frac{1}{L}\int_{-L/2}^{L/2}S(x)dx,
\end{equation}
where $\gamma_0$, $\alpha_0$ denote the averaged current density and
loss factor, respectively:
\begin{equation}
\label{ga}
\gamma_0=\frac{1}{L}\int_{-L/2}^{L/2}\gamma(x)dx,
\qquad
\alpha_0=\frac{1}{L}\int_{-L/2}^{L/2}\alpha(x)dx.
\end{equation}

It follows from the relation (\ref{sol1}) that the time-averaged
value of $\phi_t$, (being proportional to the constant voltage) is
equal to the frequency $\Omega$. Thus, the expression (\ref{iv}) can
be regarded as the current-voltage ($I$-$V$) characteristic of the
Josephson junction. One can see that the total current density
$\gamma_0$ consists of a linear (Ohmic) part and a nonlinear
contribution related to the Josephson current.

\subsection{Time-dependent equation}

For $\psi$ sufficiently small we can write
\begin{equation}
\label{sin}
\sin(\phi_0+\psi)\simeq\sin\phi_0+\psi\cos\phi_0.
\end{equation}

Noting that the term $\psi\cos\phi_0$ gives no contribution of
frequency $\Omega$ and collecting all the time-dependent terms in
Eq.~(\ref{eq1}) we find
\begin{equation}
\label{psi}
\psi_{xx} - \psi_{tt} - \alpha(x)\psi_t+\beta\psi_{xxt}\simeq\sin\phi_0.
\end{equation}

Since all the terms in Eq.~(\ref{psi}) oscillate with the same
frequency $\Omega$, it is convenient to use a complex notation
\cite{mj2,mj3}
\begin{equation}
\label{psi1}
\psi(x,t) = {\rm Im}\left[\widehat{\psi}(x)e^{i\Omega t}\right],
\qquad \sin\phi_0 = {\rm Im}\left[e^{i(\theta+\Omega t)}\right],
\end{equation}
and rewrite Eq.~(\ref{psi}) as an ordinary differential equation
\begin{equation}
\label{psi2} (1+i\Omega\beta)\widehat{\psi}_{xx} + (\Omega^2-i\alpha(x)
\Omega)\widehat{\psi} = e^{i\theta},
\end{equation}
or
\begin{equation}
\label{psi3} \widehat{\psi}_{xx} + \delta^2\widehat{\psi} =
e^{i\theta}/P,
\end{equation}
where $P=1+i\Omega\beta$, $\delta^2=(\Omega^2-i\alpha(x)\Omega)/P$.

The general approximate solution of Eq.~(\ref{psi3}) can be written
in a WKB-like form \cite{mf}
\begin{eqnarray}
\label{solpsi}
 \widehat{\psi}(x) &=& \frac{1}{2iP\sqrt\delta}\left[e^{if(x)}F^-(x)
-e^{-if(x)}F^+(x)\right]\nonumber\\
&& +A\frac{e^{if(x)}}{\sqrt\delta}+ B\frac{e^{-if(x)}}{\sqrt\delta},
\end{eqnarray}
where
\begin{equation}
\label{f} F^\pm(x)=\int_{-L/2}^x\frac{e^{i(\theta(\xi)\pm
f(\xi))}d\xi} {\sqrt\delta},\quad f(x)=\int_{-L/2}^x\delta(\xi)
d\xi.
\end{equation}

The term in square brackets in Eq.~(\ref{solpsi}) corresponds to a
dense fluxon train moving unidirectionally along the junction, while
the last two terms describe two plasma waves propagating in opposite
directions.

Using the complex formalism the boundary condition (\ref{bc3}) can
be rewritten as
\begin{equation}
\label{bc4}
\widehat{\psi}_x(\pm L/2)+i\Omega\beta\widehat{\psi}_x(\pm L/2)=0
\ \Longrightarrow\ \widehat{\psi}_x(\pm L/2)=0.
\end{equation}

Thus, using Eqs.~(\ref{solpsi}) and (\ref{bc4}) one can determine
the integration constants $A$ and $B$ to be
\begin{equation}
\label{ab}
 A=B=\frac{e^{i\delta_0 L}F^-(L/2) +e^{-i\delta_0 L}F^+(L/2)}
 {4P\sin\delta_0 L},
\end{equation}
where
\begin{equation}
\label{delta0}
\delta_0=\frac{1}{L}\int_{-L/2}^{L/2}\delta(x)dx.
\end{equation}

It should be mentioned here that the solutions for $\theta$ and
$\widehat{\psi}$ (Eqs.~(\ref{soltheta}) and (\ref{solpsi}),
respectively) are not given explicitly but rather form a system of
coupled equations. Fortunately, this system can be easily solved by
a method of consecutive iterations. Indeed, starting with $S(x)=0$
one can solve Eq.~(\ref{soltheta}) for $\theta$ and next solve
Eq.~(\ref{solpsi}) for $\widehat{\psi}$. Substituting $\theta$ and
$\widehat{\psi}$ into Eq.~(\ref{s}) we obtain a new approximation
for $S(x)$ and consequently new approximations for $\theta$ and
$\widehat{\psi}$.

It is clear from Eq.~(\ref{iv}) that $S(x)=0$ corresponds simply to
the Ohmic line $\gamma_0=\alpha_0\Omega$, while consecutive
iterations yield a sequence of approximations for $S(x)$ describing
the time-independent contribution from the Josephson current.

The rate of convergence of the iterative process depends strongly on
the junction parameters. As shown in the next Section, for moderate
values of $L$, $h$, and $\alpha_0$, the first iteration yields
satisfactory results. However, for practical FF oscillators which
are usually based on very long junctions, up to a few thousands of
iterations are needed to obtain a self-consistent solution.

\subsection{Analytical approximation}

It is interesting that for the simplest, but widely used model of
uniform bias current distribution ($\gamma(x)={\rm const}$), all the
integrations can be performed analytically, leading to a compact and
fully analytical expression for the $I$-$V$ characteristic.

Indeed, assuming the small-amplitude limit ($\widehat{\psi}\ll 1$)
and ignoring self-pumping effects ($\alpha(x)=\rm const$), one can
calculate the first-order approximation for Eq.~(\ref{iv}).
Following the steps outlined above we find $\theta(x)=hx$ and
consequently
\begin{eqnarray}
\label{psi4}
\widehat{\psi}(x) &=& -\frac{e^{ihx}}{P(h^2-\delta^2)}
+\frac{h\sin[(h+\delta)L/2]}{P(h^2-\delta^2)\delta\sin\delta L}
e^{i\delta x}\nonumber\\
&& +\frac{h\sin[(h-\delta)L/2]}{P(h^2-\delta^2)\delta\sin\delta L}
e^{-i\delta x}
\end{eqnarray}
and
\begin{equation}
\label{iv1}
\gamma_0=\alpha_0\Omega+{\rm Im}\left[-\frac{1}{2P(h^2-\delta^2)}
+\frac{h^2(\cos\delta L-\cos hL)}{LP(h^2-\delta^2)^2\delta\sin\delta L}
\right],
\end{equation}
where $P$ and $\delta$ have been defined in Eq.~(\ref{psi3}).

When the surface losses are neglected ($\beta=0$) the above simple
expression is reduced to the solution derived earlier \cite{mj2}.
Moreover, it can be shown that for $\beta=0$ the expression
(\ref{iv1}) is also equivalent to apparently different solutions
derived independently in Refs.~\onlinecite{ss2,pank1} in a form of
infinite series expansions.

\subsection{Large-amplitude corrections}

As follows from Eq.~(\ref{iv}), the evaluation of $S(x)$ is crucial
for the determination of $I$-$V$ characteristic. So far we have
assumed $\psi\ll 1$ and used an approximation (\ref{sin}). However,
for $\psi$ larger we should consider the exact relation
\begin{equation}
\label{s1}
S(x)=\langle\sin\phi_0\cos\psi\rangle_T+\langle\cos\phi_0\sin\psi\rangle_T.
\end{equation}

Since $\psi$ is an oscillatory function, one can use well-known
relations \cite{as} involving Bessel function. As shown in
Ref.~\onlinecite{mj3}, the leading terms for $\cos\psi$ and
$\sin\psi$ are given by
\begin{equation}
\label{bessel}
\cos\psi\simeq J_0(|\widehat{\psi}|),\qquad
\sin\psi\simeq\frac{2J_1(|\widehat{\psi}|)}
{|\widehat{\psi}|}\psi,
\end{equation}
where $J_0$, $J_1$ denote the Bessel functions of order 0 and 1,
respectively.

Using Eq.~(\ref{bessel}) we can find the time-independent
contribution $S(x)$ to be \cite{mj3}
\begin{equation}
\label{s2}
S(x)=\frac{J_1(|\widehat{\psi}|)}{|\widehat{\psi}|}{\rm
Im}\left[\widehat{\psi}e^{-i\theta}\right].
\end{equation}
Accordingly, the right-hand sides of Eqs.~(\ref{psi}) and
(\ref{psi2}) should be replaced by $\sin\phi_0\cos\psi$  and
$J_0(|\widehat{\psi}|)e^{i\theta}$, respectively.

\subsection{Self-pumping effects}

According to Refs. \onlinecite{pank2,pank3}, the total current
density due to the quasi-particle tunneling is given by
\begin{equation}
\label{geff1}
\gamma_{\rm eff}=\sum_{-\infty}^\infty J_n^2
\left(ev_{\rm ac}/\hbar\omega\right)\gamma_{\rm dc} (v_{\rm
dc}+n\hbar\omega/e),
\end{equation}
where $J_n$ is the Bessel functions of order $n$, $\gamma_{\rm dc}$
denotes the unpumped $I$-$V$ dependence, and the total voltage
applied to the junction can be separated into a constant ($v_{\rm
dc}$) part and and an oscillatory part of amplitude $v_{\rm ac}$ and
frequency $\omega$.

Using the Josephson relation \cite{bp} $v_{\rm dc}=\hbar\omega/2e$
and coming back to our dimensionless notation we find $v_{\rm
dc}=\Omega$, $v_{\rm ac}=|\widehat{\psi}|\Omega$ and
\begin{equation}
\label{geff2}
\gamma_{\rm eff}=\sum_{-\infty}^\infty J_n^2
\left(|\widehat{\psi}|/2\right)\gamma_{\rm dc}
\left[\Omega(1+2n)\right].
\end{equation}

For small arguments $z=|\widehat{\psi}|/2\ll 1$ the infinite
expansion (\ref{geff2}) can be truncated to include only quadratic
terms in $z$:
\begin{equation}
\label{geff3}
\gamma_{\rm eff}=\gamma_{\rm dc}(\Omega)+
\frac{z^2}{4}\left[\gamma_{\rm dc}(-\Omega)-2\gamma_{\rm dc}(\Omega)
+\gamma_{\rm dc}(3\Omega)\right].
\end{equation}

As an unpumped $I$-$V$ characteristic we can take the nonlinear
resistive model \cite{lik}
\begin{equation}
\label{gdc}
\gamma_{\rm dc}(\Omega)=\alpha_0\Omega\left\{b\frac{(\Omega/\Omega_g)^n}
{[(\Omega/\Omega_g)^n+1]}+1\right\},
\end{equation}
where $n\gg 1$, $\Omega_g$ denotes the normalized gap voltage and
$b=R_j/R_n$ is the ratio of normal-state resistances below and above
$\Omega_g$.

For $n\rightarrow\infty$ the highly nonlinear continuous dependence
(\ref{gdc}) tends to a simple discontinuous linear form which will
be used in further calculations:
\begin{equation}
\label{gdc1} \gamma_{\rm dc}=\left\{
\begin{array}{l}
\alpha_0\Omega\\
\alpha_0\Omega(b+1)
\end{array}\right.
\quad\mbox{for}\quad
\begin{array}{l}
\Omega<\Omega_g,\\
\Omega>\Omega_g.
\end{array}
\end{equation}

It is clear that $\gamma_{\rm eff}$ is $x$-dependent due to the
self-pumping via $\widehat{\psi}(x)$, thus we can define an
effective damping factor $\alpha_{\rm eff}(x)=\gamma_{\rm
eff}(x)/\Omega$ and compute it self-consistently by starting with
$\alpha_{\rm eff}=\alpha_0$, calculating $\widehat{\psi}$,
substituting into Eq.~(\ref{geff2}), evaluating a new approximation
for $\alpha_{\rm eff}(x)$, and so on.

\section{Results and discussion}

In this section we compare analytical results derived above with
numerical simulations obtained by the finite-difference implicit
scheme \cite{dodd}. The $I$-$V$ characteristic is given by
Eq.~(\ref{iv}), while $S(x)$ follows from the self-consistent
solutions of Eqs.~(\ref{soltheta}) and (\ref{solpsi}). To illustrate
the influence of surface losses we consider first the simplest case
$\gamma(x)={\rm const}$ where the $I$-$V$ dependence can be
expressed in a closed form (\ref{iv1}).

\begin{figure}
\includegraphics[scale=0.48]{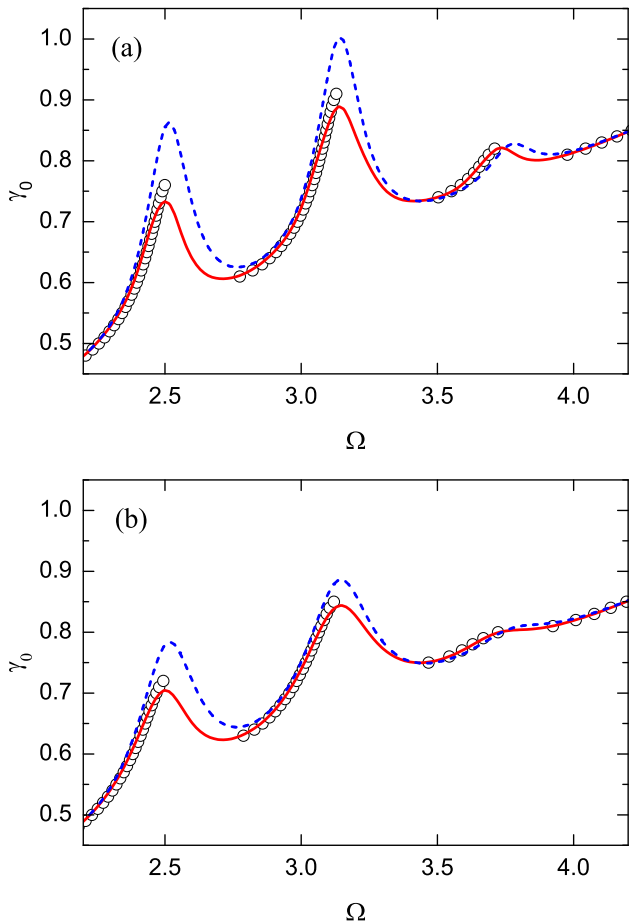}
\caption{(Color online) Current-voltage characteristic calculated
analytically (solid line) and numerically (open circles) for
$\gamma(x)=\rm const$, $L=5$, $h=3$, $\alpha_0=0.2$: (a) $\beta=0$,
(b) $\beta=0.01$. The dashed line shows the first-order
approximation (\ref{iv1}).}
\end{figure}

Fig.~1 shows the central part of the $I$-$V$ characteristic for a
junction of moderate length $L=5$. The remaining parameters
($\alpha_0=0.2$, $h=3$) are similar to those assumed in
Ref.~\onlinecite{pank1}. Open circles denote the results of
numerical simulations for a discrete set of $\gamma_0$ points. The
dashed line denotes the analytical solution (\ref{iv1}) while the
solid line follows from the self-consistent solutions
(\ref{soltheta}) and (\ref{solpsi}) obtained by consecutive
iterations for $\theta$ and $\widehat{\psi}$. Fig.~1(a) shows the
case $\beta=0$ (no surface losses) and in Fig.~1(b) we assume
$\beta=0.01$. Similar results for $h=5$ are presented in
Fig.~2(a),(b).

\begin{figure}
\includegraphics[scale=0.48]{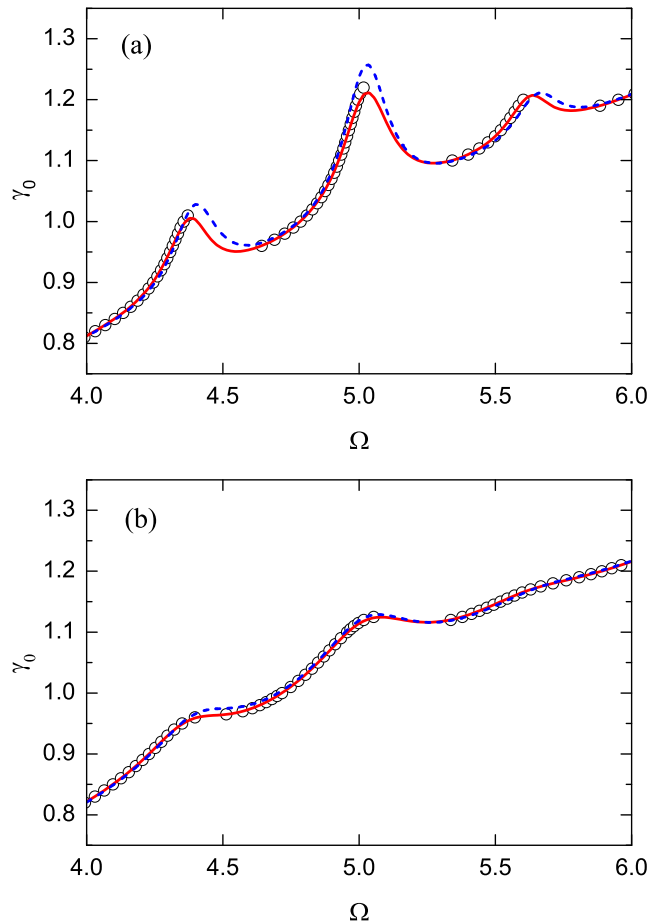}
\caption{(Color online) the same as in Fig.~1 but for $h=5$.}
\end{figure}

One can see that the $I$-$V$ dependence departs from the Ohmic line
$\gamma_0=\alpha_0\Omega$ in the region of $\Omega\simeq h$, forming
the main FF step modulated by a series of Fiske steps with voltage
spacing $\Delta\Omega\simeq\pi/L$. Analytical approximations are
continuous and consist of a series of resonances while numerical
results show typical hysteretic behavior and we observe only the
segments of positive slope.

It is clear that the presence of even very small surface losses
changes significantly the $I$-$V$ dependence. Generally, the steps
(resonances) become smaller and this effect is more pronounced for
larger values of the external field $h$. Such a result can be easily
explained if we recall that the main FF step and accompanying Fiske
steps are visible in the region $\Omega\simeq h$ and the surface
loss factor enters the formalism via $P=1+i\Omega\beta$. In other
words, for higher external magnetic field the influence of surface
losses is stronger and the $I$-$V$ characteristic becomes more
smooth. The influence of surface losses is clearly visible in
experimental results \cite{pank2,pank3}, where the Fiske steps
gradually disappear as the external magnetic field is getting
stronger.

Comparing analytical and numerical results shown in Figs.~1 and 2
one can see that the fully analytical solution (dashed line) is
fairly accurate. On the other hand, the self-consistent solution
(solid line) reproduces very accurately all the details of the
numerical solution.

As the next example, let us consider a more realistic case of a very
long junction with small damping. Following Ref.~\onlinecite{pank3}
we choose $L=40$, $\alpha_0=0.033$, $\beta=0.035$, and a slightly
asymmetric current profile $\gamma(x)$ depicted in Fig.~3. The bias
electrode ($x_1\le x\le x_2$) is shorter than the total junction
length. Consequently, the current distribution is assumed parabolic
for $x_1\le x\le x_2$, and exponential in the unbiased tails ($x\le
x_1$, $x\ge x_2$).

\begin{figure}
\includegraphics[scale=0.36]{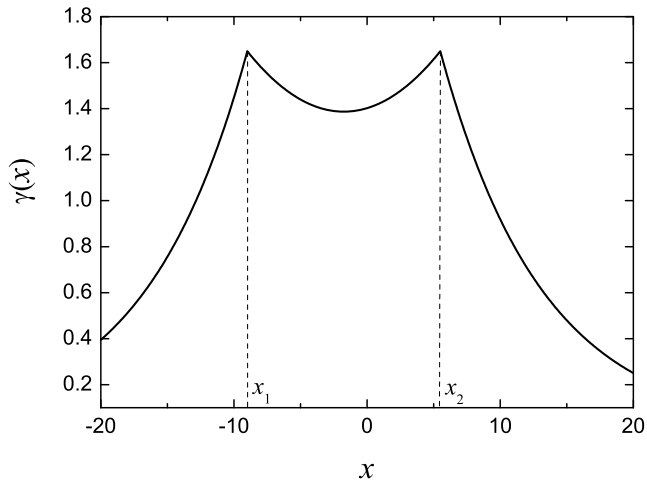}
\caption{Normalized bias current distribution similar to that
assumed in Ref. \onlinecite{pank3}. $L=40$, $x_1=-9$, $x_2=5.5$.}
\end{figure}

Contrary to the previous example, now we cannot use the analytical
approximation (\ref{iv1}). First, the current distribution is
$x$-dependent, what means that $\theta(x)$ departs from a simple
linear dependence $\theta(x)=hx$, and relevant integrals cannot be
calculated analytically. Second, it appears (even for a uniform
current profile) that the first-order approximation is not accurate
enough for very long junctions, and consecutive iterations are
necessary to obtain a self-consistent solution.

\begin{figure}
\includegraphics[scale=0.37]{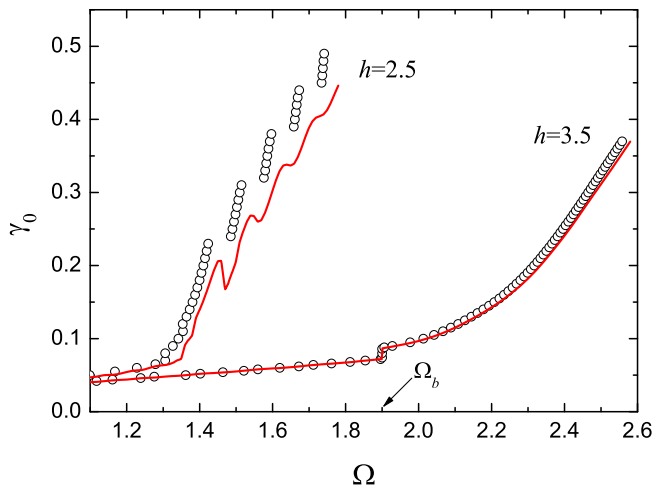}
\caption{(Color online) Current-voltage characteristic calculated
analytically (solid line) and numerically (open circles) for two
values of the external magnetic field: $h=2.5$ and $h=3.5$. The
current profile is shown in Fig.~3 and the remaining junction
parameters are: $L=40$, $\alpha_0=0.033$, $\beta=0.035$.}
\end{figure}

Fig.~4 shows the $I$-$V$ characteristics calculated for junction
parameters specified above and for two values of the external
magnetic field $h=2.5$ and $h=3.5$. As before, open circles
correspond to numerical simulations while the solid line represents
a self-consistent solution. One can see that the agreement between
numerical and analytical results is rather poor for $h=2.5$. To
explain this discrepancy we should recall the main assumption (see
Sec.~II) that Eq.~(\ref{eq1}) can be separated into a
time-independent part and a part sinusoidal in $\Omega t$ while
neglecting higher-order harmonics. The Fourier analysis shows,
however, that the anharmonic contribution to $\psi(t)$ is rather
large for $h=2.5$. E.g. for $\gamma_0=0.3$ we find the content of
the second harmonic to be about 35\%. Physically, it means that the
fluxon train is not sufficiently dense for $h=2.5$ and its
time-dependence although periodic is not strictly sinusoidal.

Contrary to the case $h=2.5$, for $h=3.5$ we observe excellent
agreement between numerical and analytical data. Now the $I$-$V$
characteristic is smooth and the Fiske steps disappeared completely
as a combined result of surface losses and self-pumping effects. The
Fourier analysis of $\psi(t)$ shows now that the anharmonicity
diminishes rather quickly with increasing $h$ and for $h=3.5$ we
observe only about 10\% of the second harmonic.

Another interesting feature which is visible for $h=3.5$ is an
additional small step at $\Omega_b=1.9$. Such a step appears in
experimental results \cite{pank3,kosh7} and can be attributed to the
self-pumping effect described in Subsection III.E.  As shown in Ref.
\onlinecite{kosh7}, the position of the current step follows from a
simple relation $\Omega_b=\Omega_g/3$, where $\Omega_g$ denotes the
gap voltage. Assuming typical junction parameters \cite{kosh7} we
find a normalized dimensionless gap voltage $\Omega_g=5.7$, hence
$\Omega_b=1.9$.

It is interesting that the value of $\Omega_b$ can also be deduced
directly form Eq.~(\ref{geff3}). Indeed, noting that $\gamma_{\rm
dc}(-\Omega)=-\gamma_{\rm dc}(\Omega)$ one can see the quadratic
term in Eq.~(\ref{geff3}) to vanish, provided $\gamma_{\rm
dc}(\Omega)$ is a linear function. However, for a nonlinearity
starting at $\Omega=\Omega_g$ we find $\gamma_{\rm dc}(3\Omega)\neq
3\gamma_{\rm dc}(\Omega)$, thus a nonzero contribution to $
\gamma_{\rm eff}$ appears at $3\Omega=\Omega_g$, giving rise to a
step at $\Omega_b=\Omega_g/3$.

\begin{figure}
\includegraphics[scale=0.36]{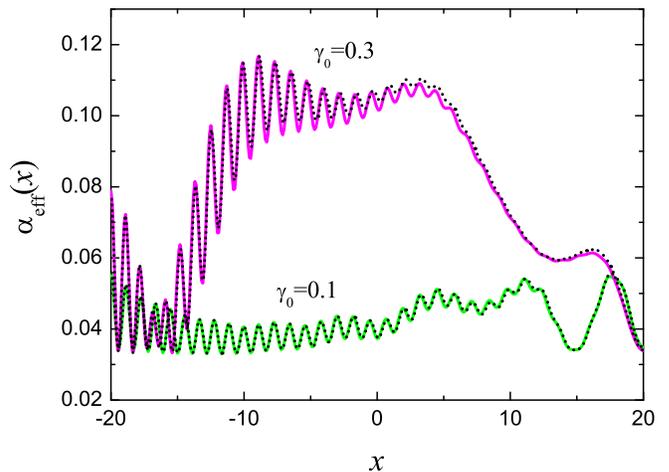}
\caption{(Color online) Spatial distribution of $\alpha_{\rm
eff}(x)$ calculated analytically (solid line) and numerically
(dotted line) for $h=3.5$ and two values of the averaged current
density $\gamma_0=0.1$ and $\gamma_0=0.3$. The remaining junction
parameters are the same as in Fig.~4.}
\end{figure}

As shown in Subsection III.E, an effective loss factor $\alpha_{\rm
eff}(x)$ is spatially modulated due to the self-pumping, and its
average value becomes significantly larger than $\alpha_0$. Fig. 5
shows the spatial distribution of $\alpha_{\rm eff}$ plotted for
$\gamma_0=0.1$ and $\gamma_0=0.3$. The junction parameters are
assumed as before. i.e. $L=40$, $h=3.5$, $\alpha_0=0.033$,
$\beta=0.035$. Now the dotted lines correspond to numerical
simulations and the solid lines denote self-consistent (analytical)
solutions. One can see that the agreement is excellent, the
self-consistent solution following closely all the details of the
numerical simulations, taken here as a reference.

It is clear that $\alpha_{\rm eff}$ generally grows with increasing
value of $\gamma_0$. E.g. for $\gamma_0=0.3$ we obtain an averaged
value of $\alpha_{\rm eff}(x)$ about three times larger than the
``unpumped'' value $\alpha_0$. Such an effect together with surface
losses makes the $I$-$V$ curve smooth, damping effectively the Fiske
steps.

\section{Conclusions}

In this paper a semi-analytical approach has been suggested, making
possible to solve a modified sG equation (\ref{sg2}) with both
surface losses and self-pumping effects taken into account. The
solution, as given by Eqs.~(\ref{sol1}), (\ref{soltheta}), and
 (\ref{solpsi}), consists of a rotating background and a
unidirectional fluxon train accompanied by two plasma waves
traveling in opposite directions with the velocity close to the
critical value $\Omega/\delta_0\simeq\pm 1$. Having obtained an
analytical solution to Eq.~(\ref{sg2}) one can determine some
macroscopic, directly measurable quantities, such as the constant
bias current and constant voltage, making possible to calculate the
$I$-$V$ characteristic (\ref{iv}) of the junction.

For the uniform current density distribution ($\gamma(x)=\rm const$)
it has been shown that the relevant expressions for $\theta(x)$,
$\widehat{\psi}(x)$, and consequently the $I$-$V$ dependence can be
obtained in a closed fully analytical form (\ref{iv1}). As follows
from Figs.~1 and 2, such an approximation is fairly accurate for
moderate junction parameters. In general, however, practical FF
oscillators are based on very long junctions with strongly
nonuniform current profile. In such a case an analytical
approximation (\ref{iv1}) appears insufficient and one should look
for a self-consistent solution of Eqs.~(\ref{soltheta}) and
(\ref{solpsi}) which can be obtained by an iterative procedure.

As shown in Figs.~1,2,4 and 5, the self-consistent solutions and
numerical simulations are generally in very good agreement. The only
exception, where one can observe qualitative rather than
quantitative agreement is the $I$-$V$ curve shown in Fig.~4 for
$h=2.5$. As explained in the previous Section, the fluxon train is
not dense enough for $h=2.5$ and consequently the time-dependent
part of the solution is periodic but not strictly sinusoidal,
violating the main assumption of the present approach.

Finally, it should be mentioned that the method presented here can
be easily extended to include general, more realistic boundary
conditions discussed in Refs. \onlinecite{pank2,pank3,sor}. In the
present paper, however, we intentionally assumed standard
open-circuit boundary conditions (\ref{bc1}), where the plasma waves
could interfere after complete reflection at the boundaries, giving
rise to clearly visible Fiske steps. This way, we were able to
separate the influence of surface losses and self-pumping from that
of boundary conditions, which (e.g. for a resistive load) can also
affect the $I$-$V$ characteristic and make the physical mechanism
discussed here less clear.

\section*{Acknowledgments}

Financial support from the Institute of Physics, Polish Academy od
Sciences is gratefully acknowledged.

\end{document}